%% file: CFT-09-019_temp.tex
\pdfoutput=1
\documentclass[11pt,twoside,a4paper,pdftex,cmspaper,final,collab]{cms-tdr}

\begin{document}\cmsNoteHeader{CFT-09-019}
%%%%%%%%%%%%%%%%%%%%%%%%%%%%%%%%%%%%%%%%%%%%%%%%%%%%%%%%%%%%%%%%%%%%
%
%  Common definitions
%
%  N.B. use of \providecommand rather than \newcommand means
%       that a definition is ignored if already specified
%
%                                              L. Taylor 18 Feb 2005
%%%%%%%%%%%%%%%%%%%%%%%%%%%%%%%%%%%%%%%%%%%%%%%%%%%%%%%%%%%%%%%%%%%%

%%%%%%%%%%%%%%%%%%%%%%%%%%%%%%%%%%%%%%%%%%%%%%%%%%%%%%%%%%%%%%%%%%%%
%
% Hyphenations (only need to add here if you get a nasty word break)
%
\hyphenation{env-iron-men-tal}%    just an example
\hyphenation{had-ron-i-za-tion}
\hyphenation{cal-or-i-me-ter}
\hyphenation{de-vices}
%
% Hyphenations-end
%
% CVS info. These are modified by cvs at checkout time.
% The last version of these macros found before the maketitle will be the one on the front page,
% so only the main file is tracked.
% Edit by hand with care!
%\pdfoutput=1
\RCS$Revision: 1.31 $
\RCS$Date: 2010/01/07 20:11:49 $
\RCS$Name:  $
%%%%%%%%%%%%% ptdr definitions %%%%%%%%%%%%%%%%%%%%%
\input{ptdr-definitions}
%%%%%%%%%%%%%%%  Title page %%%%%%%%%%%%%%%%%%%%%%%%
\cmsNoteHeader{09-019}
\title{Identification and Filtering of Uncharacteristic Noise in the CMS
Hadron Calorimeter}% Force line breaks with \\

%Author is always "The CMS Collaboration" for PAS, so author, etc will be ignored
%\address[cern]{CERN}
%\address[umd]{University of Maryland}
%\address[nwu]{Northwestern University}
%\author[umd]{Jeff Temple}\author[nwu]{Mayda Velasco}
%\author[cms]{\LARGE{The CMS Collaboration}}
\author[cms]{{The CMS Collaboration}}
% please supply the date in yyyy/mm/dd format. Today has been
% redefined to do so, but it should be fixed as of the final release date.
\date{\today}

% note that you cannot use \verb in the abstract text
\abstract{
Commissioning studies of the CMS hadron calorimeter have identified
sporadic uncharacteristic noise and a small number of malfunctioning calorimeter
channels. Algorithms have been developed to identify and address these
problems in the data.  The methods have been tested on cosmic
ray muon data, calorimeter noise data, and single beam data
collected with CMS in 2008.  The noise rejection algorithms can be
applied to LHC collision data at the trigger level or in the offline
analysis.  The application of the algorithms at the trigger level is
shown to remove 90\% of noise events with fake missing transverse
energy above 100~GeV, which is sufficient for the CMS physics trigger
operation.
}

% these need to be filled in by hand and should (MUST) match the info
% in the TeX equivalents less the TeX markup

% All authors listed here, but only editors displayed for this version
% of the paper
\hypersetup{%
pdfauthor={Jeff Temple, 
Anton Anastassov, 
John Paul Chou, 
Shuichi Kunori,
Edward Laird, 
Radek Ofierzynski, 
Kenneth Rossato, 
Mayda Velasco 
Efe Yazgan
},%
pdftitle={Identification and Filtering of Uncharacteristic Noise in the CMS
Hadron Calorimeter},%
pdfsubject={CMS},%
pdfkeywords={CMS, calorimetry, HCAL, noise, CRAFT}}

\maketitle %maketitle comes after all the front information has been supplied

%%%%%%%%%%%%%%%%%%%%%%%%%%%%%%%%  Begin text %%%%%%%%%%%%%%%%%%%%%%%%%%%%%
\section{Introduction}
\label{sec:Intro}
\input{Intro}

\section{CMS Calorimeter Description}
\label{sec:Detector}
\input{Calorimeter}

\section{Classification of HCAL Readout Problems}
\label{sec:Problem}
\input{Problems}

\section{Identification Algorithms and Performance}
\subsection{Identification}
\label{sec:IDProblems}
\input{Identification}

\subsection{Algorithm Performance}
\label{sec:Performance}
\input{Performance}

\section{Summary}

Sporadic uncharacteristic noise has been observed within the CMS hadron calorimeter during \emph{in-situ} commissioning.
A number of algorithms to identify the channels in which such noise occurs have
been designed and implemented.  These algorithms address noise and
readout issues seen in cosmic ray muon and single beam data from 2008. 

The observed noise rate is less than 8~(3)~Hz in events containing at least 45~(100)~GeV of measured MET.  This rate is reduced by an order of magnitude by excluding events flagged
by the noise algorithms as containing problematic channels.  The
algorithms have already been useful for identifying problematic
channels during CRAFT, leading to a reduction in the rate of events with large measured MET in 2009 commissioning studies.  
 
Further suppression may be gained by the application of more stringent noise
rejection algorithms.  Such algorithms reduce the noise rate to less
than 0.1~Hz for events with MET greater than 100~GeV, comparable to the expected rate from 10~TeV collisions at a luminosity of $8\times10^{29}$~cm$^{-2}$~s$^{-1}$, while reducing
the detection efficiency for various physics signals by a few percent.
These algorithms may be applied at the trigger level to
reduce noise rates, or within the offline reconstruction to separate potential noise events for further study.
Additional information for identifying noise is available from other
detectors during offline reconstruction.  This information allows for more
refined noise rejection algorithms that incur a smaller reduction in
signal efficiency.

Channels identified as problematic may be corrected during
event reconstruction or excluded from the reconstruction entirely.
Initial tests on data show that the algorithms for identifying
problematic channels correctly mark both persistent and sporadic known
problems with high efficiency.  Further tests on simulated events indicate a low mis-tag rate for properly behaving
channels.

\section*{Acknowledgments}

We thank the technical and administrative staff at CERN and other CMS
Institutes, and acknowledge support from: FMSR (Austria); FNRS and FWO
(Belgium); CNPq, CAPES, FAPERJ, and FAPESP (Brazil); MES (Bulgaria);
CERN; CAS, MoST, and NSFC (China); COLCIENCIAS (Colombia); MSES
(Croatia); RPF (Cyprus); Academy of Sciences and NICPB (Estonia);
Academy of Finland, ME, and HIP (Finland); CEA and CNRS/IN2P3
(France); BMBF, DFG, and HGF (Germany); GSRT (Greece); OTKA and NKTH
(Hungary); DAE and DST (India); IPM (Iran); SFI (Ireland); INFN
(Italy); NRF (Korea); LAS (Lithuania); CINVESTAV, CONACYT, SEP, and
UASLP-FAI (Mexico); PAEC (Pakistan); SCSR (Poland); FCT (Portugal);
JINR (Armenia, Belarus, Georgia, Ukraine, Uzbekistan); MST and MAE
(Russia); MSTDS (Serbia); MICINN and CPAN (Spain); Swiss Funding
Agencies (Switzerland); NSC (Taipei); TUBITAK and TAEK (Turkey); STFC
(United Kingdom); DOE and NSF (USA). Individuals have received support
from the Marie-Curie IEF program (European Union); the Leventis
Foundation; the A. P. Sloan Foundation; and the Alexander von Humboldt
Foundation.  

\vfill\eject

\bibliography{auto_generated}   % will be created by the tdr script.
\cleardoublepage\appendix\section{The CMS Collaboration \label{app:collab}}\begin{sloppypar}\hyphenpenalty=500\input{CFT-09-019-authorlist.tex}\end{sloppypar}
\end{document}

%% file: ptdr-definitions.tex
%%%%%%%%%%%%%%%%%%%%%%%%%%%%%%%%%%%%%%%%%%%%%%%%%%%%%%%%%%%%%%%%%%%%
%
%  Common definitions
%
%  N.B. use of \providecommand rather than \newcommand means
%       that a definition is ignored if already specified
%
%                                              L. Taylor 18 Feb 2005
%%%%%%%%%%%%%%%%%%%%%%%%%%%%%%%%%%%%%%%%%%%%%%%%%%%%%%%%%%%%%%%%%%%%

% Some shorthand
% turn off italics
\newcommand {\etal}{\mbox{et al.}\xspace} %et al. - no preceding comma
\newcommand {\ie}{\mbox{i.e.}\xspace}     %i.e.
\newcommand {\eg}{\mbox{e.g.}\xspace}     %e.g.
\newcommand {\etc}{\mbox{etc.}\xspace}     %etc.
\newcommand {\vs}{\mbox{\sl vs.}\xspace}      %vs.
\newcommand {\mdash}{\ensuremath{\mathrm{-}}} % for use within formulas

% some terms whose definition we may change
\newcommand {\Lone}{Level-1\xspace} % Level-1 or L1 ?
\newcommand {\Ltwo}{Level-2\xspace}
\newcommand {\Lthree}{Level-3\xspace}

% Some software programs (alphabetized)
\providecommand{\ACERMC} {\textsc{AcerMC}\xspace}
\providecommand{\ALPGEN} {{\textsc{alpgen}}\xspace}
\providecommand{\CHARYBDIS} {{\textsc{charybdis}}\xspace}
\providecommand{\CMKIN} {\textsc{cmkin}\xspace}
\providecommand{\CMSIM} {{\textsc{cmsim}}\xspace}
\providecommand{\CMSSW} {{\textsc{cmssw}}\xspace}
\providecommand{\COBRA} {{\textsc{cobra}}\xspace}
\providecommand{\COCOA} {{\textsc{cocoa}}\xspace}
\providecommand{\COMPHEP} {\textsc{CompHEP}\xspace}
\providecommand{\EVTGEN} {{\textsc{evtgen}}\xspace}
\providecommand{\FAMOS} {{\textsc{famos}}\xspace}
\providecommand{\GARCON} {\textsc{garcon}\xspace}
\providecommand{\GARFIELD} {{\textsc{garfield}}\xspace}
\providecommand{\GEANE} {{\textsc{geane}}\xspace}
\providecommand{\GEANTfour} {{\textsc{geant4}}\xspace}
\providecommand{\GEANTthree} {{\textsc{geant3}}\xspace}
\providecommand{\GEANT} {{\textsc{geant}}\xspace}
\providecommand{\HDECAY} {\textsc{hdecay}\xspace}
\providecommand{\HERWIG} {{\textsc{herwig}}\xspace}
\providecommand{\HIGLU} {{\textsc{higlu}}\xspace}
\providecommand{\HIJING} {{\textsc{hijing}}\xspace}
\providecommand{\IGUANA} {\textsc{iguana}\xspace}
\providecommand{\ISAJET} {{\textsc{isajet}}\xspace}
\providecommand{\ISAPYTHIA} {{\textsc{isapythia}}\xspace}
\providecommand{\ISASUGRA} {{\textsc{isasugra}}\xspace}
\providecommand{\ISASUSY} {{\textsc{isasusy}}\xspace}
\providecommand{\ISAWIG} {{\textsc{isawig}}\xspace}
\providecommand{\MADGRAPH} {\textsc{MadGraph}\xspace}
\providecommand{\MCATNLO} {\textsc{mc@nlo}\xspace}
\providecommand{\MCFM} {\textsc{mcfm}\xspace}
\providecommand{\MILLEPEDE} {{\textsc{millepede}}\xspace}
\providecommand{\ORCA} {{\textsc{orca}}\xspace}
\providecommand{\OSCAR} {{\textsc{oscar}}\xspace}
\providecommand{\PHOTOS} {\textsc{photos}\xspace}
\providecommand{\PROSPINO} {\textsc{prospino}\xspace}
\providecommand{\PYTHIA} {{\textsc{pythia}}\xspace}
\providecommand{\SHERPA} {{\textsc{sherpa}}\xspace}
\providecommand{\TAUOLA} {\textsc{tauola}\xspace}
\providecommand{\TOPREX} {\textsc{TopReX}\xspace}
\providecommand{\XDAQ} {{\textsc{xdaq}}\xspace}

%  Experiments
\newcommand {\DZERO}{D\O\xspace}     %etc.

% Measurements and units...

\newcommand{\de}{\ensuremath{^\circ}}
\newcommand{\ten}[1]{\ensuremath{\times \text{10}^\text{#1}}}
\newcommand{\unit}[1]{\ensuremath{\text{\,#1}}\xspace}
\newcommand{\mum}{\ensuremath{\,\mu\text{m}}\xspace}
\newcommand{\micron}{\ensuremath{\,\mu\text{m}}\xspace}
\newcommand{\cm}{\ensuremath{\,\text{cm}}\xspace}
\newcommand{\mm}{\ensuremath{\,\text{mm}}\xspace}
\newcommand{\mus}{\ensuremath{\,\mu\text{s}}\xspace}
\newcommand{\keV}{\ensuremath{\,\text{ke\hspace{-.08em}V}}\xspace}
\newcommand{\MeV}{\ensuremath{\,\text{Me\hspace{-.08em}V}}\xspace}
\newcommand{\GeV}{\ensuremath{\,\text{Ge\hspace{-.08em}V}}\xspace}
\newcommand{\TeV}{\ensuremath{\,\text{Te\hspace{-.08em}V}}\xspace}
\newcommand{\PeV}{\ensuremath{\,\text{Pe\hspace{-.08em}V}}\xspace}
\newcommand{\keVc}{\ensuremath{{\,\text{ke\hspace{-.08em}V\hspace{-0.16em}/\hspace{-0.08em}c}}}\xspace}
\newcommand{\MeVc}{\ensuremath{{\,\text{Me\hspace{-.08em}V\hspace{-0.16em}/\hspace{-0.08em}c}}}\xspace}
\newcommand{\GeVc}{\ensuremath{{\,\text{Ge\hspace{-.08em}V\hspace{-0.16em}/\hspace{-0.08em}c}}}\xspace}
\newcommand{\TeVc}{\ensuremath{{\,\text{Te\hspace{-.08em}V\hspace{-0.16em}/\hspace{-0.08em}c}}}\xspace}
\newcommand{\keVcc}{\ensuremath{{\,\text{ke\hspace{-.08em}V\hspace{-0.16em}/\hspace{-0.08em}c}^\text{2}}}\xspace}
\newcommand{\MeVcc}{\ensuremath{{\,\text{Me\hspace{-.08em}V\hspace{-0.16em}/\hspace{-0.08em}c}^\text{2}}}\xspace}
\newcommand{\GeVcc}{\ensuremath{{\,\text{Ge\hspace{-.08em}V\hspace{-0.16em}/\hspace{-0.08em}c}^\text{2}}}\xspace}
\newcommand{\TeVcc}{\ensuremath{{\,\text{Te\hspace{-.08em}V\hspace{-0.16em}/\hspace{-0.08em}c}^\text{2}}}\xspace}

\newcommand{\pbinv} {\mbox{\ensuremath{\,\text{pb}^\text{$-$1}}}\xspace}
\newcommand{\fbinv} {\mbox{\ensuremath{\,\text{fb}^\text{$-$1}}}\xspace}
\newcommand{\nbinv} {\mbox{\ensuremath{\,\text{nb}^\text{$-$1}}}\xspace}
\newcommand{\percms}{\ensuremath{\,\text{cm}^\text{$-$2}\,\text{s}^\text{$-$1}}\xspace}
\newcommand{\lumi}{\ensuremath{\mathcal{L}}\xspace}
\newcommand{\Lumi}{\ensuremath{\mathcal{L}}\xspace}%both upper and lower
%
% Need a convention here:
\newcommand{\LvLow}  {\ensuremath{\mathcal{L}=\text{10}^\text{32}\,\text{cm}^\text{$-$2}\,\text{s}^\text{$-$1}}\xspace}
\newcommand{\LLow}   {\ensuremath{\mathcal{L}=\text{10}^\text{33}\,\text{cm}^\text{$-$2}\,\text{s}^\text{$-$1}}\xspace}
\newcommand{\lowlumi}{\ensuremath{\mathcal{L}=\text{2}\times \text{10}^\text{33}\,\text{cm}^\text{$-$2}\,\text{s}^\text{$-$1}}\xspace}
\newcommand{\LMed}   {\ensuremath{\mathcal{L}=\text{2}\times \text{10}^\text{33}\,\text{cm}^\text{$-$2}\,\text{s}^\text{$-$1}}\xspace}
\newcommand{\LHigh}  {\ensuremath{\mathcal{L}=\text{10}^\text{34}\,\text{cm}^\text{$-$2}\,\text{s}^\text{$-$1}}\xspace}
\newcommand{\hilumi} {\ensuremath{\mathcal{L}=\text{10}^\text{34}\,\text{cm}^\text{$-$2}\,\text{s}^\text{$-$1}}\xspace}

% Some usual physics terms

\newcommand{\zp}{\ensuremath{\mathrm{Z}^\prime}\xspace}

% SM (still to be classified)

\newcommand{\kt}{\ensuremath{k_{\mathrm{T}}}\xspace}
\newcommand{\BC}{\ensuremath{{B_{\mathrm{c}}}}\xspace}
\newcommand{\bbarc}{\ensuremath{{\overline{b}c}}\xspace}
\newcommand{\bbbar}{\ensuremath{{b\overline{b}}}\xspace}
\newcommand{\ccbar}{\ensuremath{{c\overline{c}}}\xspace}
\newcommand{\JPsi}{\ensuremath{{J}/\psi}\xspace}
\newcommand{\bspsiphi}{\ensuremath{B_s \to \JPsi\, \phi}\xspace}
\newcommand{\AFB}{\ensuremath{A_\mathrm{FB}}\xspace}
\newcommand{\EE}{\ensuremath{e^+e^-}\xspace}
\newcommand{\MM}{\ensuremath{\mu^+\mu^-}\xspace}
\newcommand{\TT}{\ensuremath{\tau^+\tau^-}\xspace}
\newcommand{\wangle}{\ensuremath{\sin^{2}\theta_{\mathrm{eff}}^\mathrm{lept}(M^2_\mathrm{Z})}\xspace}
\newcommand{\ttbar}{\ensuremath{{t\overline{t}}}\xspace}
\newcommand{\stat}{\ensuremath{\,\text{(stat.)}}\xspace}
\newcommand{\syst}{\ensuremath{\,\text{(syst.)}}\xspace}
% these moved to similar defs
%\newcommand{\Etmiss}{\ensuremath{E_{\mathrm{T}\!{\rm miss}}}}
%\newcommand{\VEtmiss}{\ensuremath{{\vec E}_{\mathrm{T}\!{\rm miss}}}}

%%%  E-gamma definitions
\newcommand{\HGG}{\ensuremath{\mathrm{H}\to\gamma\gamma}}
\newcommand{\gev}{\GeV}
\newcommand{\GAMJET}{\ensuremath{\gamma + \mathrm{jet}}}
\newcommand{\PPTOJETS}{\ensuremath{\mathrm{pp}\to\mathrm{jets}}}
\newcommand{\PPTOGG}{\ensuremath{\mathrm{pp}\to\gamma\gamma}}
\newcommand{\PPTOGAMJET}{\ensuremath{\mathrm{pp}\to\gamma +
\mathrm{jet}
}}
\newcommand{\MH}{\ensuremath{\mathrm{M_{\mathrm{H}}}}}
\newcommand{\RNINE}{\ensuremath{\mathrm{R}_\mathrm{9}}}
\newcommand{\DR}{\ensuremath{\Delta\mathrm{R}}}

% Physics symbols ...

\newcommand{\PT}{\ensuremath{p_{\mathrm{T}}}\xspace}
\newcommand{\pt}{\ensuremath{p_{\mathrm{T}}}\xspace}
\newcommand{\ET}{\ensuremath{E_{\mathrm{T}}}\xspace}
\newcommand{\HT}{\ensuremath{H_{\mathrm{T}}}\xspace}
\newcommand{\et}{\ensuremath{E_{\mathrm{T}}}\xspace}
\newcommand{\Em}{\ensuremath{E\!\!\!/}\xspace}
\newcommand{\Pm}{\ensuremath{p\!\!\!/}\xspace}
\newcommand{\PTm}{\ensuremath{{p\!\!\!/}_{\mathrm{T}}}\xspace}
\newcommand{\ETm}{\ensuremath{E_{\mathrm{T}}^{\mathrm{miss}}}\xspace}
\newcommand{\MET}{\ensuremath{E_{\mathrm{T}}^{\mathrm{miss}}}\xspace}
\newcommand{\ETmiss}{\ensuremath{E_{\mathrm{T}}^{\mathrm{miss}}}\xspace}
\newcommand{\VEtmiss}{\ensuremath{{\vec E}_{\mathrm{T}}^{\mathrm{miss}}}\xspace}

%%%%%%
% From Albert
%

\newcommand{\ga}{\ensuremath{\gtrsim}}
\newcommand{\la}{\ensuremath{\lesssim}}
\newcommand{\swsq}{\ensuremath{\sin^2\theta_W}\xspace}
\newcommand{\cwsq}{\ensuremath{\cos^2\theta_W}\xspace}
\newcommand{\tanb}{\ensuremath{\tan\beta}\xspace}
\newcommand{\tanbsq}{\ensuremath{\tan^{2}\beta}\xspace}
\newcommand{\sidb}{\ensuremath{\sin 2\beta}\xspace}
\newcommand{\alpS}{\ensuremath{\alpha_S}\xspace}
\newcommand{\alpt}{\ensuremath{\tilde{\alpha}}\xspace}

\newcommand{\QL}{\ensuremath{Q_L}\xspace}
\newcommand{\sQ}{\ensuremath{\tilde{Q}}\xspace}
\newcommand{\sQL}{\ensuremath{\tilde{Q}_L}\xspace}
\newcommand{\ULC}{\ensuremath{U_L^C}\xspace}
\newcommand{\sUC}{\ensuremath{\tilde{U}^C}\xspace}
\newcommand{\sULC}{\ensuremath{\tilde{U}_L^C}\xspace}
\newcommand{\DLC}{\ensuremath{D_L^C}\xspace}
\newcommand{\sDC}{\ensuremath{\tilde{D}^C}\xspace}
\newcommand{\sDLC}{\ensuremath{\tilde{D}_L^C}\xspace}
\newcommand{\LL}{\ensuremath{L_L}\xspace}
\newcommand{\sL}{\ensuremath{\tilde{L}}\xspace}
\newcommand{\sLL}{\ensuremath{\tilde{L}_L}\xspace}
\newcommand{\ELC}{\ensuremath{E_L^C}\xspace}
\newcommand{\sEC}{\ensuremath{\tilde{E}^C}\xspace}
\newcommand{\sELC}{\ensuremath{\tilde{E}_L^C}\xspace}
\newcommand{\sEL}{\ensuremath{\tilde{E}_L}\xspace}
\newcommand{\sER}{\ensuremath{\tilde{E}_R}\xspace}
\newcommand{\sFer}{\ensuremath{\tilde{f}}\xspace}
\newcommand{\sQua}{\ensuremath{\tilde{q}}\xspace}
\newcommand{\sUp}{\ensuremath{\tilde{u}}\xspace}
\newcommand{\suL}{\ensuremath{\tilde{u}_L}\xspace}
\newcommand{\suR}{\ensuremath{\tilde{u}_R}\xspace}
\newcommand{\sDw}{\ensuremath{\tilde{d}}\xspace}
\newcommand{\sdL}{\ensuremath{\tilde{d}_L}\xspace}
\newcommand{\sdR}{\ensuremath{\tilde{d}_R}\xspace}
\newcommand{\sTop}{\ensuremath{\tilde{t}}\xspace}
\newcommand{\stL}{\ensuremath{\tilde{t}_L}\xspace}
\newcommand{\stR}{\ensuremath{\tilde{t}_R}\xspace}
\newcommand{\stone}{\ensuremath{\tilde{t}_1}\xspace}
\newcommand{\sttwo}{\ensuremath{\tilde{t}_2}\xspace}
\newcommand{\sBot}{\ensuremath{\tilde{b}}\xspace}
\newcommand{\sbL}{\ensuremath{\tilde{b}_L}\xspace}
\newcommand{\sbR}{\ensuremath{\tilde{b}_R}\xspace}
\newcommand{\sbone}{\ensuremath{\tilde{b}_1}\xspace}
\newcommand{\sbtwo}{\ensuremath{\tilde{b}_2}\xspace}
\newcommand{\sLep}{\ensuremath{\tilde{l}}\xspace}
\newcommand{\sLepC}{\ensuremath{\tilde{l}^C}\xspace}
\newcommand{\sEl}{\ensuremath{\tilde{e}}\xspace}
\newcommand{\sElC}{\ensuremath{\tilde{e}^C}\xspace}
\newcommand{\seL}{\ensuremath{\tilde{e}_L}\xspace}
\newcommand{\seR}{\ensuremath{\tilde{e}_R}\xspace}
\newcommand{\snL}{\ensuremath{\tilde{\nu}_L}\xspace}
\newcommand{\sMu}{\ensuremath{\tilde{\mu}}\xspace}
\newcommand{\sNu}{\ensuremath{\tilde{\nu}}\xspace}
\newcommand{\sTau}{\ensuremath{\tilde{\tau}}\xspace}
\newcommand{\Glu}{\ensuremath{g}\xspace}
\newcommand{\sGlu}{\ensuremath{\tilde{g}}\xspace}
\newcommand{\Wpm}{\ensuremath{W^{\pm}}\xspace}
\newcommand{\sWpm}{\ensuremath{\tilde{W}^{\pm}}\xspace}
\newcommand{\Wz}{\ensuremath{W^{0}}\xspace}
\newcommand{\sWz}{\ensuremath{\tilde{W}^{0}}\xspace}
\newcommand{\sWino}{\ensuremath{\tilde{W}}\xspace}
\newcommand{\Bz}{\ensuremath{B^{0}}\xspace}
\newcommand{\sBz}{\ensuremath{\tilde{B}^{0}}\xspace}
\newcommand{\sBino}{\ensuremath{\tilde{B}}\xspace}
\newcommand{\Zz}{\ensuremath{Z^{0}}\xspace}
\newcommand{\sZino}{\ensuremath{\tilde{Z}^{0}}\xspace}
\newcommand{\sGam}{\ensuremath{\tilde{\gamma}}\xspace}
\newcommand{\chiz}{\ensuremath{\tilde{\chi}^{0}}\xspace}
\newcommand{\chip}{\ensuremath{\tilde{\chi}^{+}}\xspace}
\newcommand{\chim}{\ensuremath{\tilde{\chi}^{-}}\xspace}
\newcommand{\chipm}{\ensuremath{\tilde{\chi}^{\pm}}\xspace}
\newcommand{\Hone}{\ensuremath{H_{d}}\xspace}
\newcommand{\sHone}{\ensuremath{\tilde{H}_{d}}\xspace}
\newcommand{\Htwo}{\ensuremath{H_{u}}\xspace}
\newcommand{\sHtwo}{\ensuremath{\tilde{H}_{u}}\xspace}
\newcommand{\sHig}{\ensuremath{\tilde{H}}\xspace}
\newcommand{\sHa}{\ensuremath{\tilde{H}_{a}}\xspace}
\newcommand{\sHb}{\ensuremath{\tilde{H}_{b}}\xspace}
\newcommand{\sHpm}{\ensuremath{\tilde{H}^{\pm}}\xspace}
\newcommand{\hz}{\ensuremath{h^{0}}\xspace}
\newcommand{\Hz}{\ensuremath{H^{0}}\xspace}
\newcommand{\Az}{\ensuremath{A^{0}}\xspace}
\newcommand{\Hpm}{\ensuremath{H^{\pm}}\xspace}
\newcommand{\sGra}{\ensuremath{\tilde{G}}\xspace}
\newcommand{\mtil}{\ensuremath{\tilde{m}}\xspace}
\newcommand{\rpv}{\ensuremath{\rlap{\kern.2em/}R}\xspace}
\newcommand{\LLE}{\ensuremath{LL\bar{E}}\xspace}
\newcommand{\LQD}{\ensuremath{LQ\bar{D}}\xspace}
\newcommand{\UDD}{\ensuremath{\overline{UDD}}\xspace}
\newcommand{\Lam}{\ensuremath{\lambda}\xspace}
\newcommand{\Lamp}{\ensuremath{\lambda'}\xspace}
\newcommand{\Lampp}{\ensuremath{\lambda''}\xspace}
\newcommand{\spinbd}[2]{\ensuremath{\bar{#1}_{\dot{#2}}}\xspace}

\newcommand{\MD}{\ensuremath{{M_\mathrm{D}}}\xspace}% ED mass
\newcommand{\Mpl}{\ensuremath{{M_\mathrm{Pl}}}\xspace}% Planck mass
\newcommand{\Rinv} {\ensuremath{{R}^{-1}}\xspace}

%%%%%%%%%%%%%%%%%%%%%%%%%%%%%%%%%%%%%%%%%%%%%%%%%%%%%%%%%%%%%%%%%%%%
%
% Hyphenations (only need to add here if you get a nasty word break)
%
\hyphenation{en-viron-men-tal}%    just an example

%% file: Intro.tex
The primary goal of the Compact Muon Solenoid (CMS) experiment \cite{CMSjinst} is to explore physics at the TeV energy scale, exploiting the proton-proton collisions delivered by the Large Hadron Collider (LHC) \cite{LHC}.  The CMS hadron calorimeter (HCAL), together with the electromagnetic
calorimeter, form a complete calorimetry system for the
measurements of jets and missing transverse energy (MET).  The measurement of jets and MET is essential for identifying many Standard
Model processes (e.g., QCD multi-jets, top, $W$+jets, and $Z$+jets) as well
as new physics signatures.  

Early commissioning studies have identified a small number of
non-responsive readout channels within the hadron calorimeter \cite{HcalNoise}.
Low rates of sporadic noise, in which the hadron
calorimeter reports spurious energy deposition, have also been
observed.  Although the overall rate is low, such noise can result in
large values of measured MET.  Triggers that rely on MET can thus be particularly affected by this
noise.  Rejection algorithms have been developed to remove events
identified as noisy at the trigger level.

The results presented here are based on data collected from detector commissioning
tests without the solenoidal field and with the solenoidal magnet
operating at its nominal axial field strength of 3.8~T,
and from initial LHC tests circulating single beams of protons.  A
collection of 300 million cosmic ray muon events was recorded from the Cosmic RUn at ZEro
Tesla (CRUZET), a set of commissioning runs taken between May and
September 2008 with the magnetic field off. A month-long detector-commissioning exercise
undertaken from October to November 2008, referred to as the Cosmic Run At
Four Tesla (CRAFT), was used to collect 270~million cosmic ray-triggered events with the CMS solenoidal magnet ramped to 3.8~T.  Data from single beams of
450~GeV/c protons were collected in September 2008, with the solenoid off.  Single beam data included
``halo'' events, in which muons produced from interactions of off-axis
beam protons with beamline elements traversed the detector, and
``splash'' events, in which the beam was intentionally directed into
closed collimators 150~m upstream of CMS, resulting in high muon fluxes and large energy
signatures in the calorimeter. 

An overview of the CMS hadron calorimeter hardware and readout
electronics chain is provided in Section \ref{sec:Detector}.  The construction of physics objects from the
hardware output is also addressed.  Observed problems in the hadron
calorimeter output discovered during detector commissioning are
described in Section \ref{sec:Problem}.  Section \ref{sec:IDProblems}
details the algorithms used to detect and flag such
problems, as well as methods for filtering events identified as noise.
Section \ref{sec:Performance} summarizes the performance of noise
identification and filtering algorithms on commissioning data and Monte Carlo (MC) simulated events.

%% file: Calorimeter.tex
The CMS detector is composed of an inner silicon tracking system
surrounded by a scintillating crystal electromagnetic calorimeter, a
brass/steel sampling hadron calorimeter, and muon detectors.  The tracker and the
majority of the calorimeters lie within the cryostat of the solenoidal magnet.  The muon detectors and portions of HCAL lie outside the
cryostat.  Data from all these systems are filtered through a two-level
trigger system. The first level (L1), composed of custom hardware
processors, uses information from the calorimeters and muon detectors
to select the most interesting events with a latency of two microseconds. The High
Level Trigger (HLT) processor farm further decreases the event rate
from 100~kHz to 100~Hz before data storage \cite{CMSjinst}.  HLT paths based on
missing transverse energy are particularly sensitive to HCAL noise, as
discussed in Section \ref{sec:IDProblems}.

\subsection{HCAL Design}
\begin{figure}[!Hhtb]
  \centering
  \includegraphics[scale=.8]{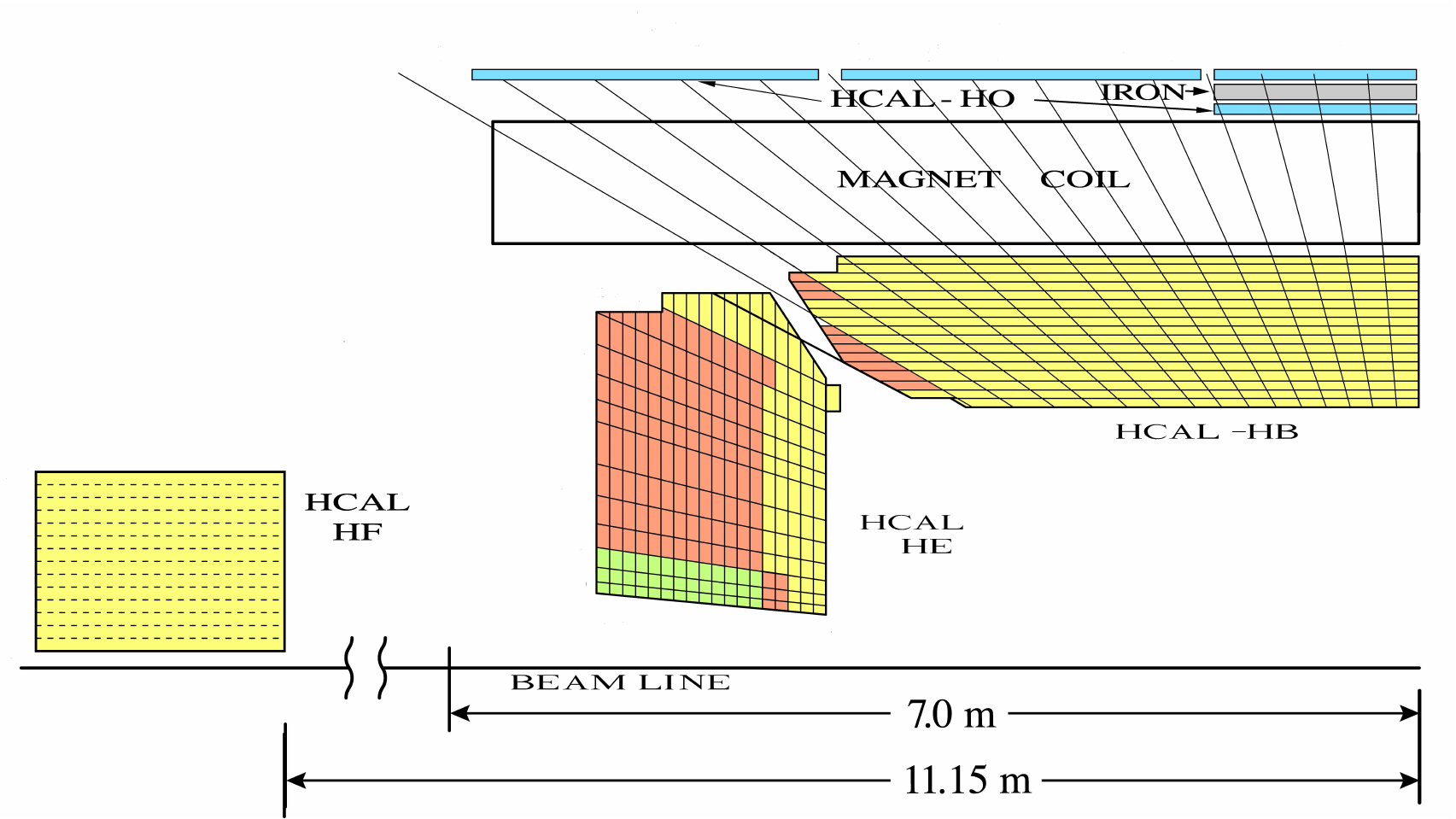}
  \caption{Quarter view of the CMS hadron calorimeter.  The shading indicates the optical grouping of scintillator layers into different longitudinal readouts.}
  \label{fig:HCAL_subdet}
\end{figure}

The HCAL detector is divided into four subdetectors (Fig.
\ref{fig:HCAL_subdet}), comprising a total of 9072 channels.  The HCAL barrel (HB) and endcap (HE)
detectors surround the electromagnetic calorimeter, and are contained
completely within the high magnetic field region of the solenoid.  HB
provides coverage in the pseudorapidity ($\eta$) range $|\eta|<1.4$, while HE provides overlapping coverage in the region $1.3<|\eta|<3.0$.  The HCAL forward calorimeters (HF) provide
measurements of energetic forward jets and increase the hermeticity of
the missing transverse energy measurement.  The HF subdetectors extend
the HCAL pseudorapidity coverage into the $|\eta|$ region 2.9--5.0. The
effective HCAL thickness in the region $|\eta|<1.3$ is extended by
the addition of an array of ``outer barrel'' (HO) scintillators
outside the magnet cryostat.  Each subdetector spans the full range of the azimuthal angle $\phi$.

The HB and HE subdetectors consist of layers of plastic scintillator
within a  brass/stainless steel absorber.  These subdetectors are segmented into
readout channels that cover an area of 0.087$\times$0.087 in $\eta-\phi$
space.  In the regions where $|\eta|$ is greater than 1.74, the
$\phi$~segmentation is more coarsely granulated.  Scintillation light
is detected by hybrid photodiodes (HPDs), with each HPD collecting
signals from 18 different HCAL channels.

The HF subdetector is a Cerenkov light detector
made of quartz fibers embedded within a 165~cm long steel absorber.
There are two types of fibers within HF:  ``long'' fibers that span the 
length of the subdetector, and ``short'' fibers that begin 22~cm into
the detector.  Differences between signals read out from the long and
short fibers are used to distinguish between electromagnetic and hadronic showers.  Long and short fibers are separately grouped to span 0.174 radians in $\phi$, and intervals in $\eta$ ranging between 0.111 and 0.178.  Each group is read out separately as a single HF channel.  (In the region $|\eta|>4.7$, each HF channel covers 0.348 rad in $\phi$.)  Photomultiplier
tubes (PMTs) connected to the fibers via light guides convert detected light to
electrical signals.

Seven-bit analog-to-digital converters (ADCs) digitize the signals from the
calorimeter for readout.  Signals from 4 HPDs or 72 PMTs are digitized
within a single 72-channel readout box (RBX).  Further discussions of
the design and performance of the HB, HE, and HF subdetectors can be
found in Refs.~\cite{HBdesign,HFdesign}.

\subsection{Reconstruction of HCAL Objects}
\label{sec:Reconstruction}

The ADC output of each calorimeter channel is computed every 25~ns, once per LHC clock cycle.  The size and shape of the ADC output over a
period of up to ten consecutive 25~ns intervals (``time
slices''), along with the appropriate conversion constants, allow the measurement of the energy deposited and the time when the energy deposition occurred.  Calibration with LED and laser sources, as well as with radioactive sources of known energy, ensures that the conversion constants provide the correct output energy values \cite{CMSjinst}. 

The CMS reconstruction software stores the reported energy and time
information for all HCAL channels with energy greater than 0.3--0.8 GeV in an event, with the exact threshold varying by subdetector and geometric location of the channel.
A ``quality flag'' is also stored for each channel.  Quality flags are used to identify specific occurrences in which otherwise
well-behaving channels display anomalous responses, as described in Section \ref{sec:IDProblems}.  The quality flags also indicate whether additional corrections were applied when reconstructing a channel's energy and time information.  Channels that
show persistent problems are tracked within a separate status
database \cite{CONDDB}.  Information from both the quality flags and the status database
is used in constructing physics objects from the HCAL channels.

Reconstructed HCAL energies are combined with information from the
electromagnetic calorimeter to form
projective calorimeter towers.  These towers contain the sums of
energies from their constituent electromagnetic calorimeter and HCAL
channels.  A channel
that is identified as problematic, either by its quality flag or by
information in the status database, may be excluded from the formation of
the calorimeter tower, or used to mark the tower as problematic.
Information about rejected or problematic channels within a calorimeter
tower is in turn used to reject or flag jet and
particle objects reconstructed from the towers.

A user-controlled set of parameters defines how the quality
flag and status database information are used during event reconstruction.  Channel quality flags are used to flag or reject problematic towers within individual events, while the channel database information is used to mark all towers containing persistent problematic channels.

%The rejection or flagging of calorimeter towers based on
%quality flags is applied only to events in which the flags in question
%are set, while rejection based on a channel's database information is
%applied to all events in which that channel is present.

%%Summary information from potentially noisy channels is also added to the data
%%stream during event reconstruction.  This information may be used for
%%analysis-specific investigations of calorimeter noise. 

%% file: Problems.tex
A number of channels with intermittent and persistent problems have been observed in the CRAFT and CRUZET commissioning data.  The classification schemes for defining each type of problem are given below. 

The following types of intermittent signals have been observed thus
far:

\begin{itemize}

\item{\textbf{Ion Feedback}.  HCAL HPDs occasionally generate
  appreciable signals even when no light is incident on their
  photo-cathodes.  Such signals are predominantly caused by a
  thermally emitted electron ionizing a gas or surface molecule in the
  acceleration gap of the HPD. That ion is accelerated back to the
  cathode and liberates further electrons, causing a signal equivalent
  to many photo-electrons.  This behavior typically manifests itself as
  a significant energy deposit in a single channel of an HPD within an event.  Ionization near an edge or corner between the hexagonal HPD pixels may lead to deposits in two or three channels.}

\item{\textbf{Other HPD Noise}.  It is known that the presence of an
  external magnetic field can alter the flashover voltage of
  dielectric materials \cite{HPDNOISE}.  Misalignments
  between the electric field within an HPD and the external solenoid
  field can lower the flashover voltage of the HPD.  This can lead to
  an avalanche of secondary electrons, producing significant energy
  deposition in a large number of channels within an HPD.  In addition,
  large energy deposits in multiple HPD channels have been observed
  even with the solenoidal field off.  Signals appearing in a large
  number of channels within a single HPD, regardless of the state of the
  magnetic field, are categorized as ``HPD noise''.}

\item{\textbf{RBX Noise}.  Events have been observed in
  which nearly all of the 72 channels within a single HB or HE RBX report
  large observed energies.  Though the cause of this is not yet understood, its distinctive signature allows it to be identified within an event.}

\item{\textbf{PMT Window Hits}.  Occasionally an energetic charged particle directly
impinges upon the window of an HF PMT rather than striking a HF quartz
fiber.  This results in an abnormally large apparent energy signal for
a signal channel within the HF.  Such a signal may be identified by
comparing the reported energies between associated long and short
fibers at a given ($\eta$, $\phi$) position. }

\item{\textbf{ADC Saturation}.  The ADC output of each channel was designed to cover
  a large dynamic range, allowing for energies up to
  1.4--2.4~TeV in each HB channel, 1.7--2.4~TeV in HE, 2.0--4.4~TeV in HF, and
  3.0~TeV in HO.  Although these limits are not expected to be
  exceeded during collision data collection,
  energies beyond these thresholds have been observed in ``beam splash'' data.}

\end{itemize}

Examples of persistent problems include channels that fail to report
correctly formatted data and channels that continuously report higher-than-expected
energies.  These are referred to as ``dead'' and ``hot'' channels,
respectively.  Dead or hot channels can
likewise result in incorrectly measured jets and biased MET measurements.  Algorithms for identifying both intermittent and
persistent problems are described in Section \ref{sec:IDProblems}.

\begin{figure}[!Hhtbp]
 \centering
 \includegraphics[scale=.55]{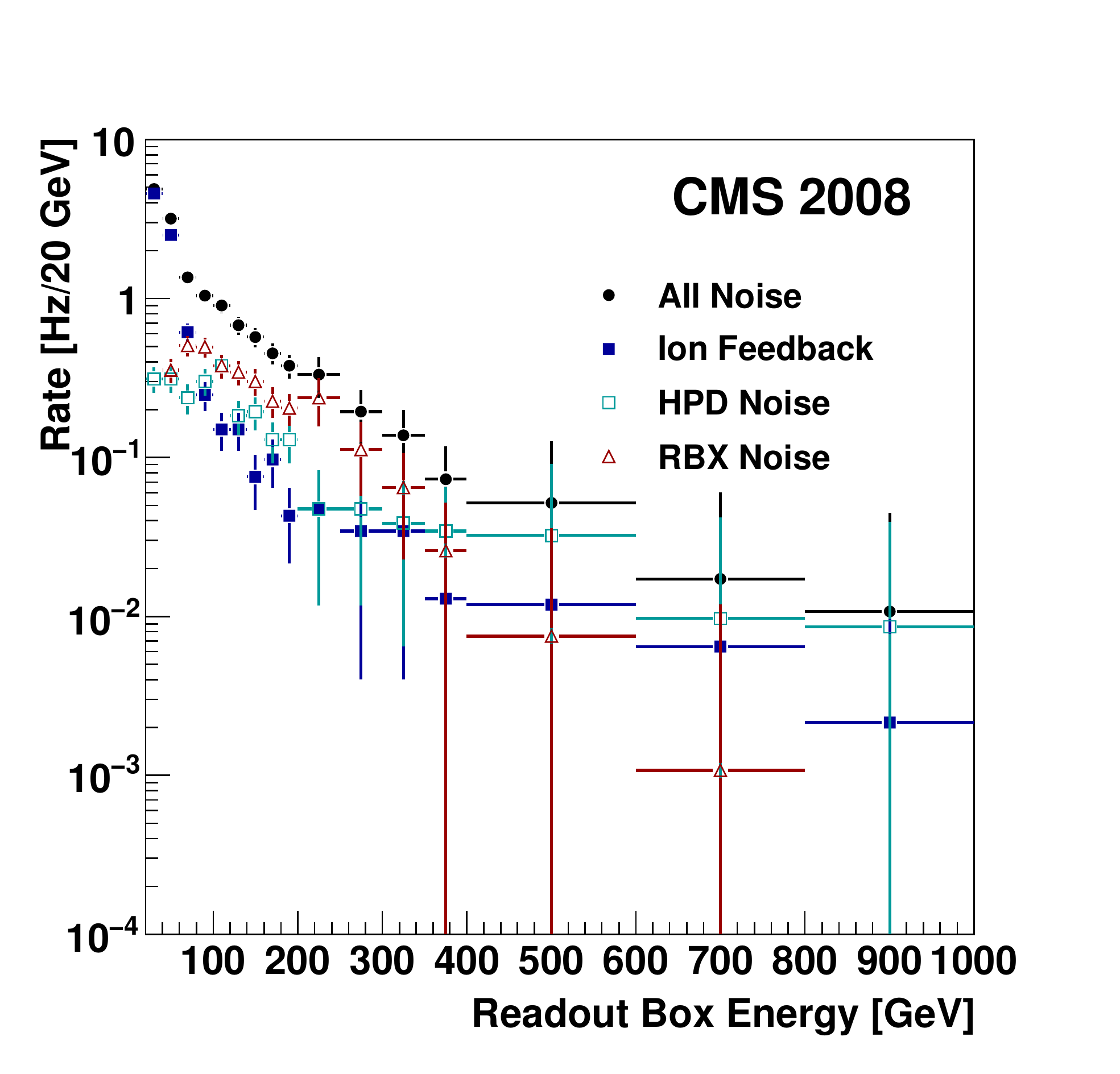}
 \includegraphics[scale=.55]{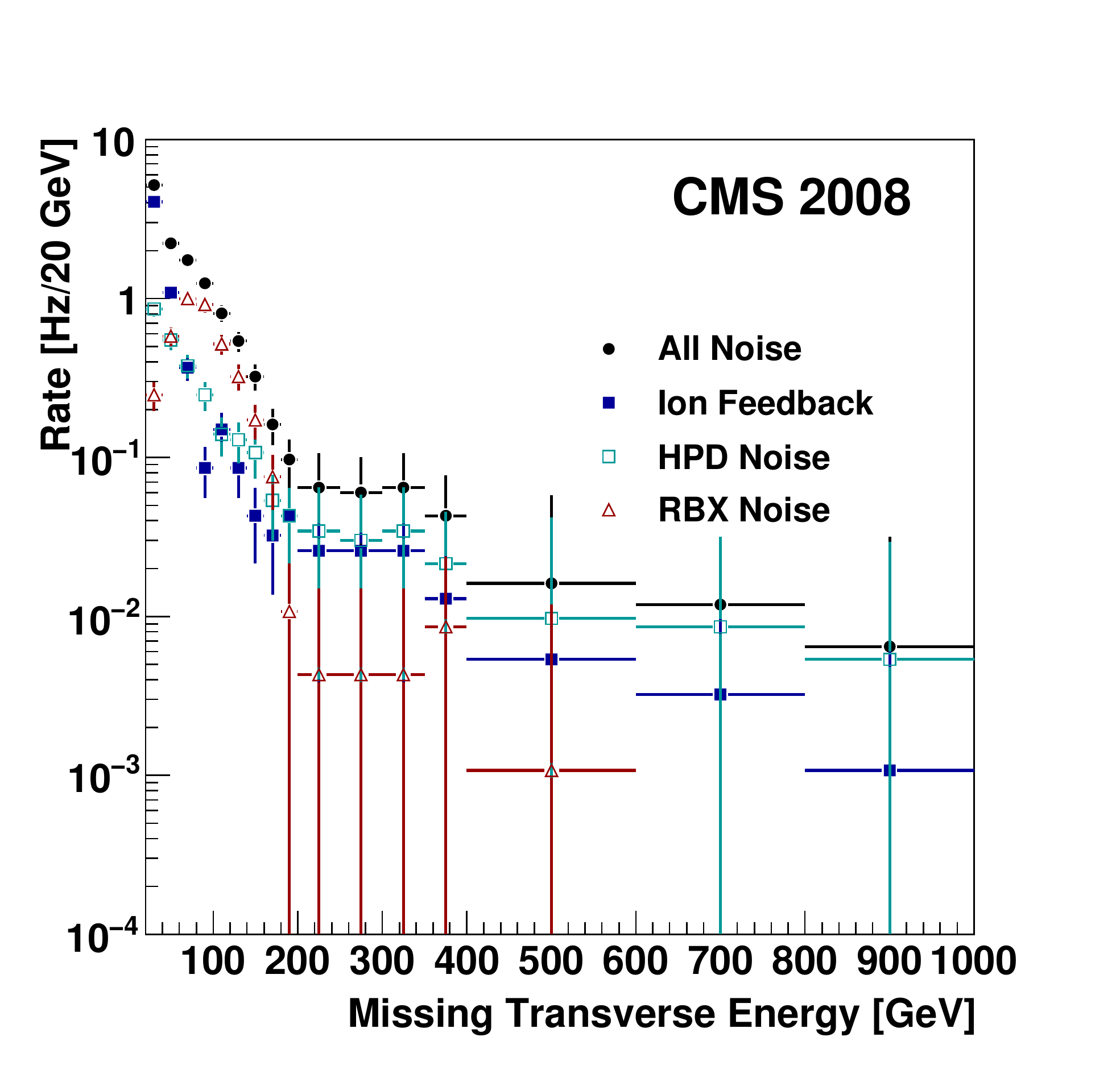}

 \caption{HCAL noise rates from CRAFT data passing a 6~GeV jet trigger.  Rates are plotted vs. (top) readout box energy and (bottom) missing transverse energy for all events with at least one HB/HE readout box containing more than 20~GeV of energy.  Only channels with energy greater than 1~GeV are considered when calculating contributions to the readout box energy.}
 \label{fig:HCAL_noiserates}
\end{figure}

Noise rates were determined from a set of 13\,000 CRAFT events
triggered on calorimeter jets with at least 6~GeV of transverse
energy.  Although attempts were made to remove events with cosmic ray muons, a small number of air shower or muon
bremsstrahlung events may remain within the sample.  Figure \ref{fig:HCAL_noiserates} shows the rates of
intermittent noise in HB and HE versus RBX energy and MET.  The MET is calculated from energy deposits in HB, HE, and the electromagnetic calorimeter.  Only RBXs that recorded at
least 20~GeV of total deposited energy were considered in computing
the noise rates.  Rates were determined from the approximately 280 (out of 288)
HPDs within HB and HE that were ramped to their nominal high voltage
during CRAFT. 

For the purposes of these rate calculations, an RBX is considered to
exhibit ``RBX noise'' in an event if at least 19 channels report energies of
1~GeV or more.  Since each HPD contains 18 channels, this requirement
ensures that the noisy behavior is not contained within a single
HPD.  An RBX containing between 10 and 18 channels with energy greater
than 1~GeV is considered to exhibit ``HPD noise'', while an RBX with one to nine
such channels is classified as exhibiting ``ion feedback''.  This last
classification encompasses both true ion feedback and signals from
cosmic ray muons.  These definitions allow for the presence of multiple
noise types within an event in the case that more than one RBX
exhibits noise.  Two of the 13\,000 examined events were found to contain more than one noisy RBX.    

\begin{figure}[!Hhtb]
  \centering
  \includegraphics[scale=.55]{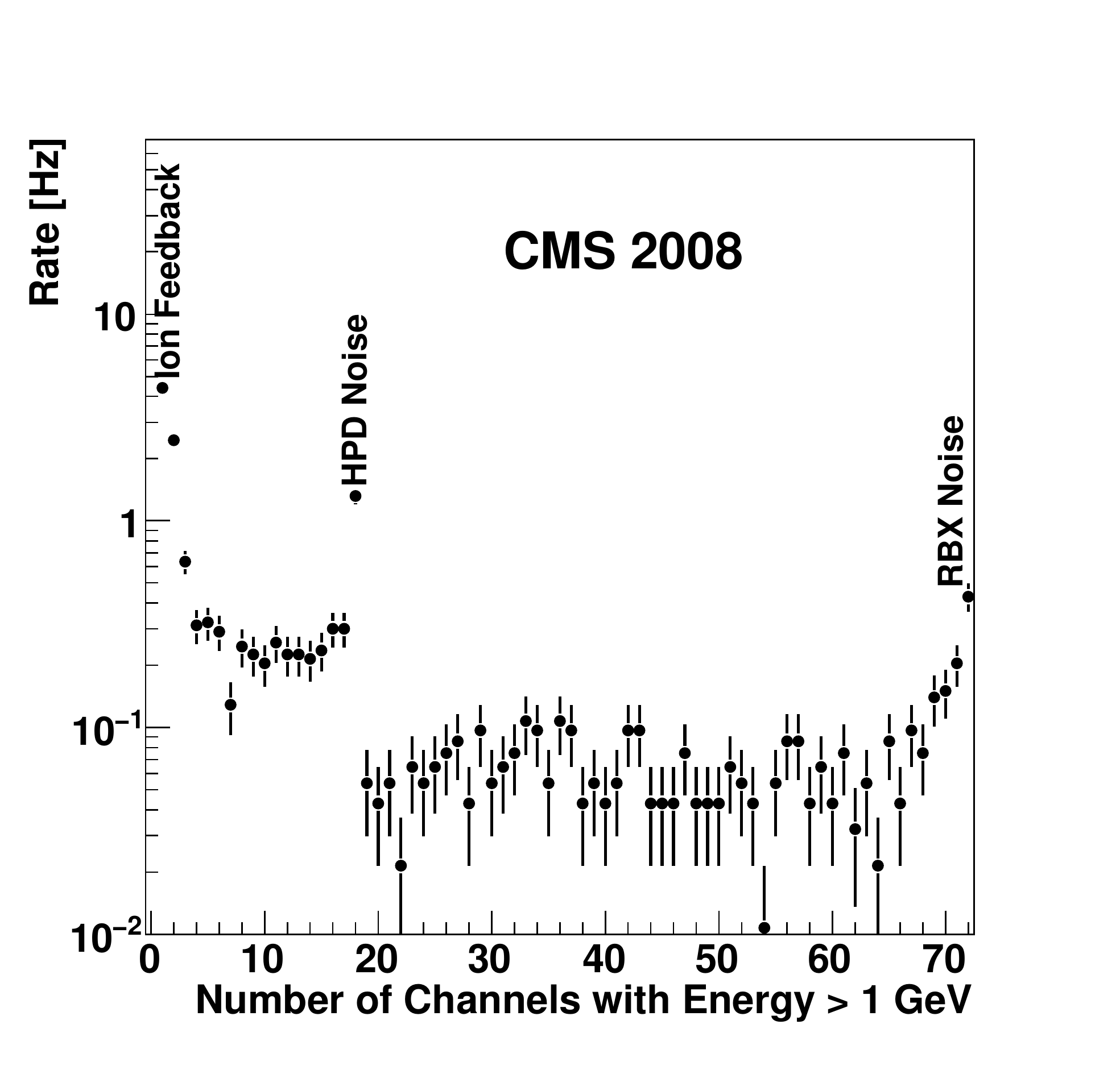}
  \caption{Number of HCAL channels within a single RBX with energy
  greater than 1~GeV, for readout boxes with a total energy greater than 20
  GeV.  Data were collected from CRAFT events triggered on calorimeter jets with at least 6~GeV of transverse energy.}
  \label{fig:HCAL_noisetypes}
\end{figure}

The distribution of the number of channels above the 1~GeV threshold
is shown in Fig. \ref{fig:HCAL_noisetypes} for RBXs with total energy
greater than 20~GeV.  Clear peaks are visible
at 18 and 72 channels.  These are the expected signatures for HPD
noise and RBX noise, respectively.

The set of triggers to be used for collecting initial collision data
at CMS is expected to include two triggers based on the MET calculated at the HLT.  These two triggers will use MET thresholds of 45~GeV and 100~GeV, respectively.  The
rates of these triggers due to physics events have been estimated from a
sample of simulated minimum bias events at a collision energy of 10~TeV.  The respective rates of the 45~GeV and 100~GeV triggers at a luminosity of $8\times10^{29}$~cm$^{-2}$~s$^{-1}$ have been estimated to be 0.3~Hz and
less than 0.1~Hz.  As shown in Fig. \ref{fig:HCAL_noiserates}, the rate of noise events
passing each MET trigger is more than an order of magnitude larger than the
rate of physics events.

%% file: Identification.tex
Algorithms within the CMS reconstruction software have been devised
for identifying intermittent HCAL problems due to HPD noise, RBX
noise, PMT window hits, and ADC saturation.  Because ion feedback typically occurs within a
single channel, and on a timescale smaller than the 25~ns clock pulse of the
accelerator, it is not 
explicitly flagged as HCAL noise.  It is instead treated as a
contribution to the uncertainty in jet energy resolution
\cite{HcalNoise}.  Separate algorithms have been developed to identify
persistently hot and dead channels.

If at least 17 of the 18 channels within an HPD report at least 1~GeV
of energy, all channels within that HPD are flagged as noisy for that
event.  This requirement flags channels that contribute to the peaks labeled as ``HPD Noise'' and ``RBX Noise'' in Fig. \ref{fig:HCAL_noisetypes}.  A second check on barrel and endcap noise is performed by evaluating the pulse shape in HB/HE signals.  A typical HB/HE signal contains a
sharp peak followed by a ``tail'' that extends for several time slices.
The tail is defined as the summed charge in the three time intervals
following the three intervals contributing to the bulk of the charge
distribution, as illustrated in Fig. 4.  A channel is flagged as noisy if the
charge contained in the tail of the distribution is outside an
expected range when compared to the total integrated charge, as shown in Fig. \ref{fig:HBHEBAD} and detailed in Section \ref{sec:Performance}.  Similar effects appear in both magnet-on and magnet-off data.

\begin{figure}[!Hhtb]
\centering

\includegraphics[scale=0.35]{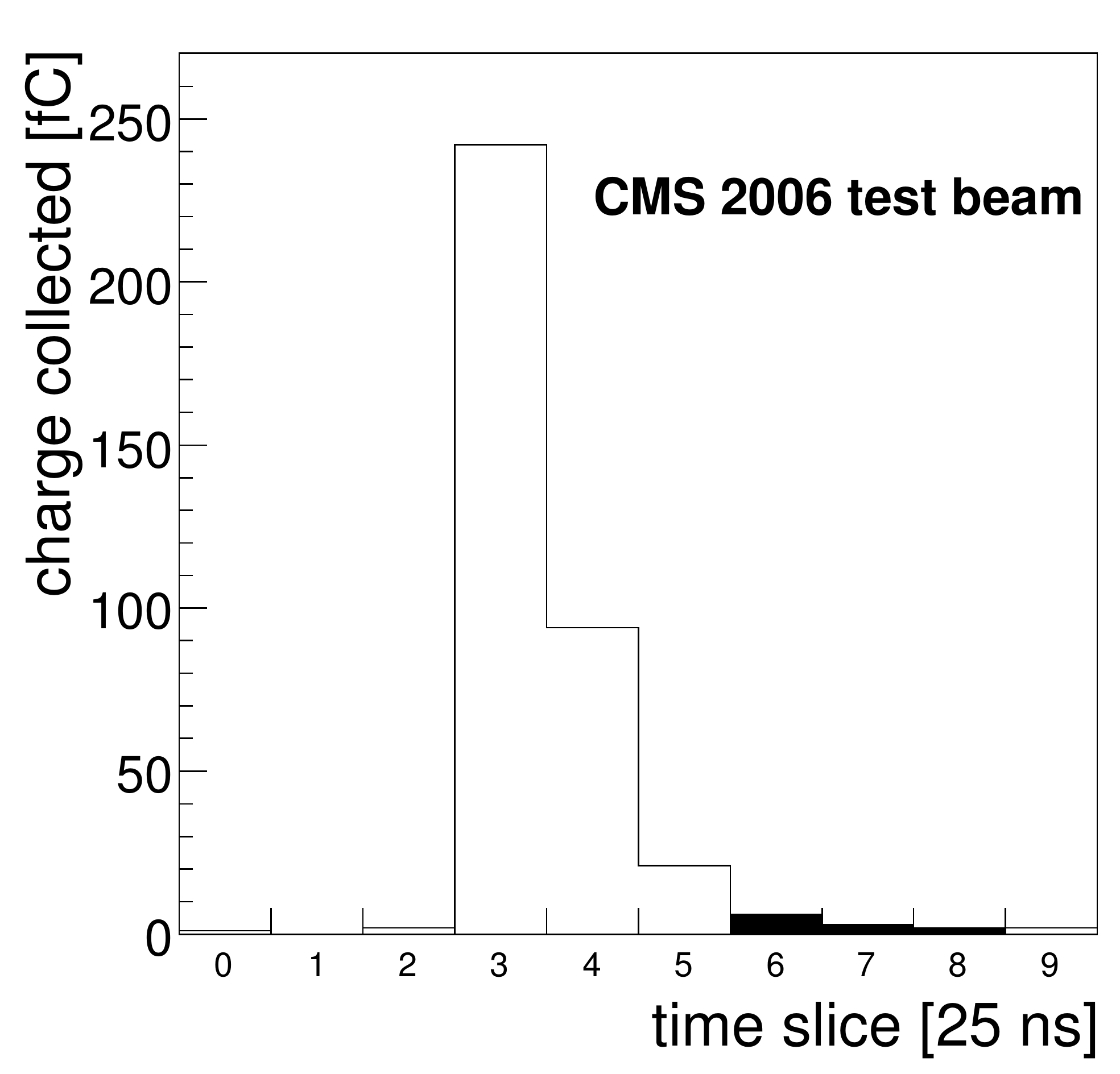}
\includegraphics[scale=0.35]{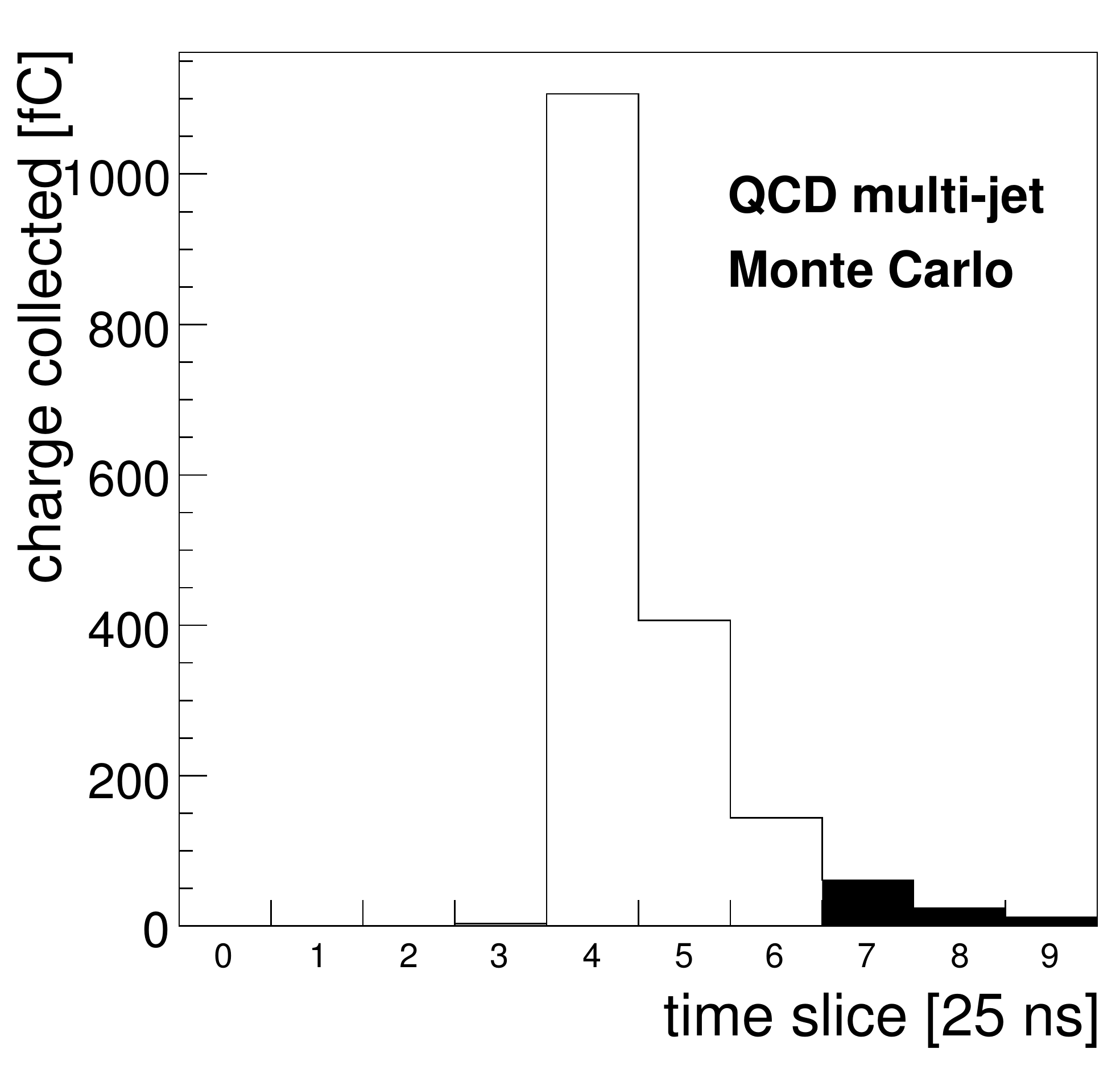}

\caption[]{Expected HCAL signal pulse shape, from
(left) pion test beam data \cite{HBdesign} and
(right) a simulated QCD multi-jet sample .  The tails of each pulse are shaded in black.}

\end{figure}

\begin{figure}[!Hhtb]
\centering
\includegraphics[scale=0.35]{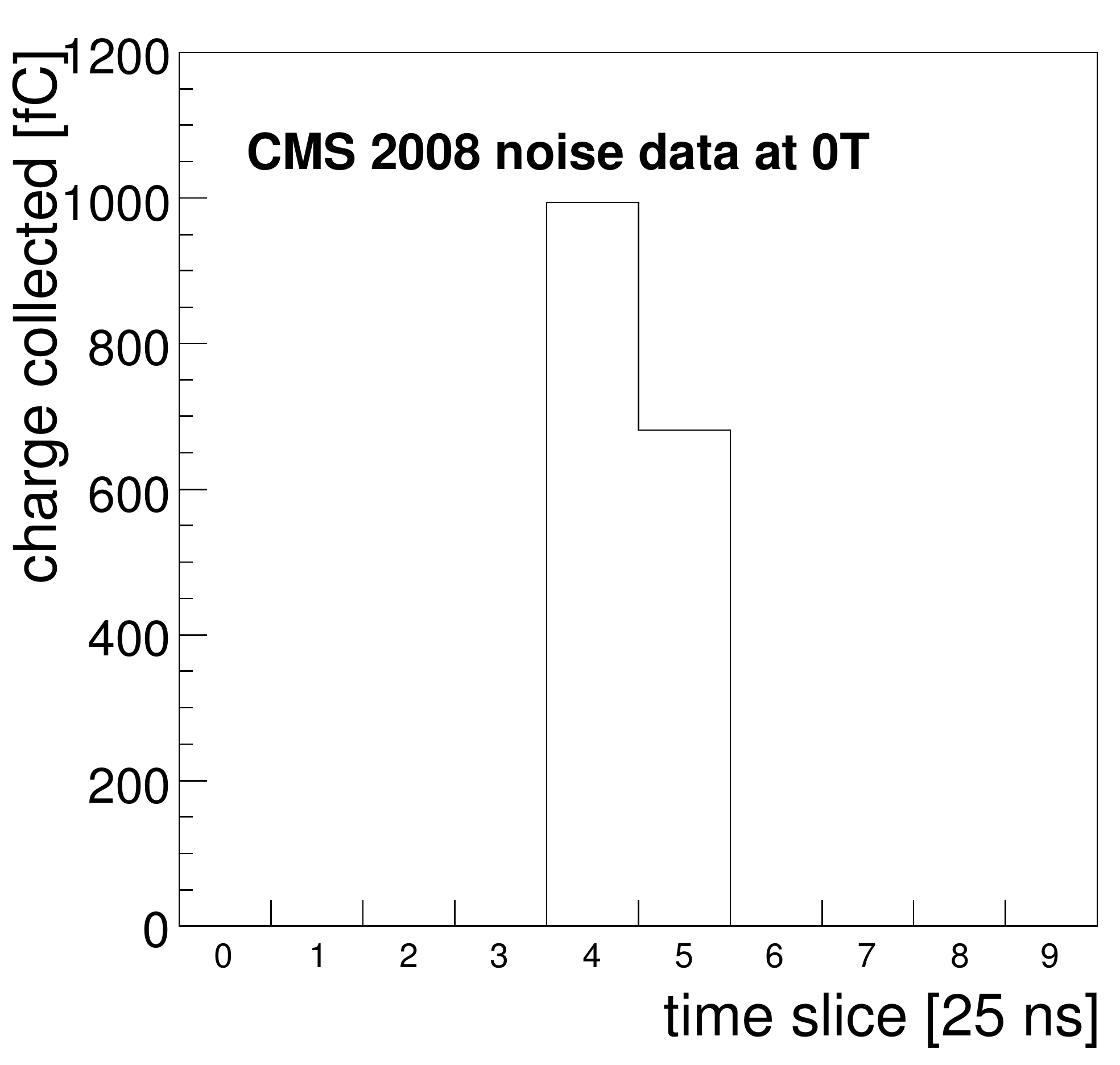}
\includegraphics[scale=0.35]{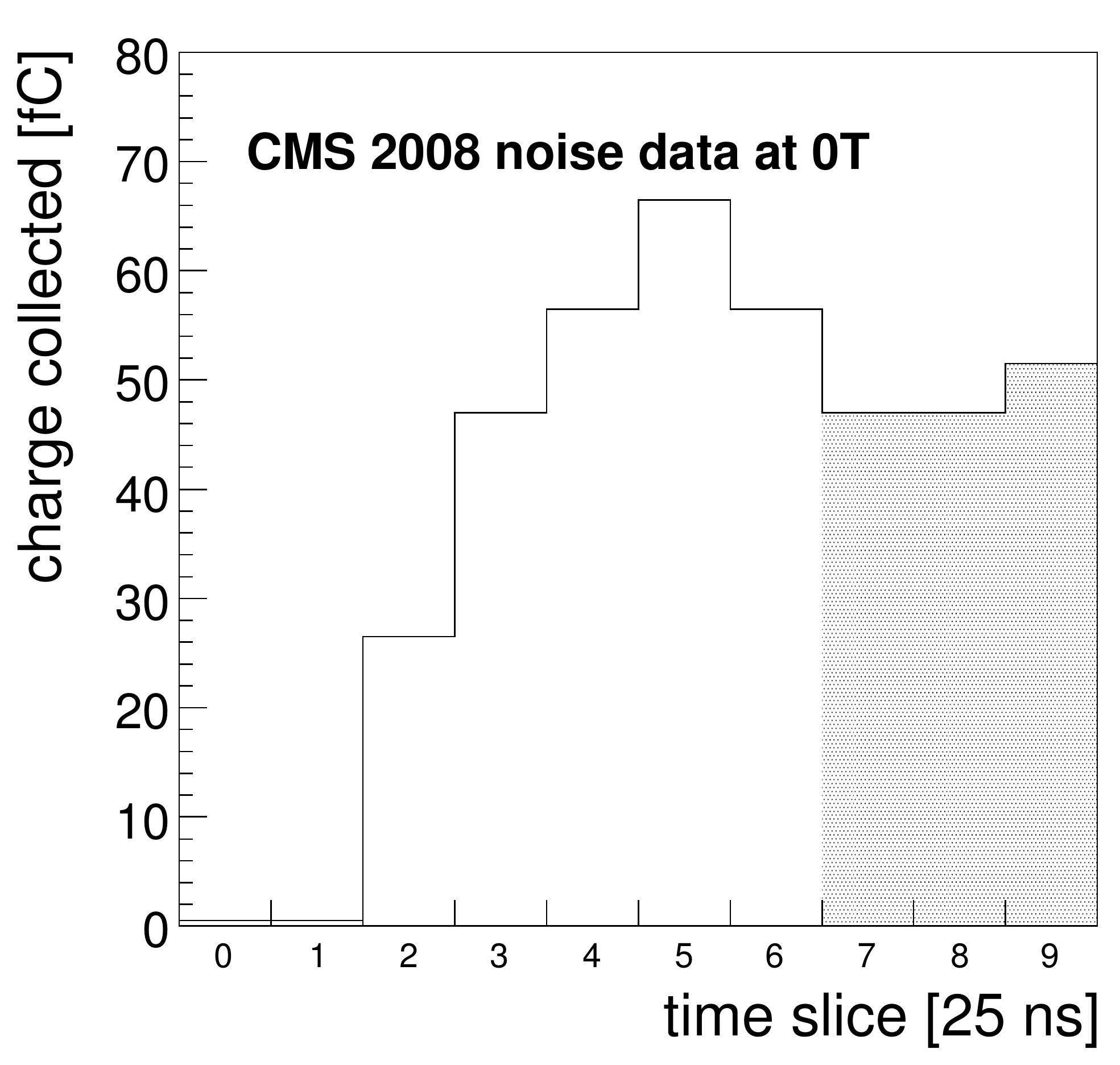}

\caption[]{ Each figure shows a charge distribution from a single HB channel in a CRUZET event.  Events were collected from a trigger that required a minimum energy of 10~GeV from any two neighboring HCAL towers.
The
channel on the left has
no tail, while the (shaded) tail of the channel on the right is abnormally large.  Both channels are flagged as
noisy. }
\label{fig:HBHEBAD}
\end{figure}

Although 
channels within a noisy HPD or RBX typically report large energies,
some channels may fall well \textit{below} the expected energy scale
associated with electronics noise during the 250~ns
readout period.  These channels report zero ADC counts during the noise event.  A large multiplicity ($\geq 8$) of such channels within an RBX is an indicator of RBX or HPD noise.

The timing of energy deposits in a channel relative to the trigger can also
indicate HPD or RBX noise.  The unusual pulse shape associated with the noise can result in a reconstructed time that is
substantially offset from the trigger decision time.  The relative
timing of a high energy deposit ($\gtrsim 25$~GeV) can be reconstructed
typically within $\pm 5$~ns, so a reconstructed time outside this window is
characteristic of HCAL noise \cite{HCAL_PHIL}.

PMT window hits within the HF are flagged by comparing the energies
reconstructed from long and short fibers with the same ($\eta,\phi$)
values.  Given a pair of adjacent long and short fibers, with energies
$E_L, E_S$, respectively, the fiber channels are flagged as noisy if 
(a) the maximum transverse energy reported by the two channels is at
least 2~GeV, and (b) if the energy ratio $R$, defined as

\begin{equation}
R\equiv\left|\frac{E_L-E_S}{E_L+E_S}\right|\quad,
\end{equation}

is greater than 0.99.

The bulk of the signal charge from an HF hit is expected to be
collected within a
single 25~ns time slice \cite{HFdesign}.  An HF channel is
flagged as noisy if (a) a significant ADC signal peak ($>10$
counts) is found, and (b) the ratio of
that peak value to the counts in the window immediately before (after)
the peak is less than 2.5 (1.0).  These constraints will be tightened
as the HF timing is refined.

Another set of HCAL quality flags is used to determine the
reliability of the energy and time information reported by each
channel.  Any channel reporting a saturated ADC level is flagged as problematic, since the correspondence between measured and true energy is lost.  Channels whose readouts are out of synchronization with the rest of the detector are also flagged as problematic.

At present, persistently problematic channels are defined as those channels
exhibiting abnormal behavior in at least 5\% of the events collected in a
given data period.  A channel is considered hot if it reports 
at least 5~GeV of energy in at least 5\% of collected events.  A channel is also
considered hot if it remains above a threshold energy of 3~GeV for at
least 1\,000 consecutive events, and if this condition persists for at
least 5\% of the collected events.  These two definitions serve
complementary purposes, with the first aiding in the identification of
hot channels energetic enough to affect trigger rates, and the second
serving as a scan for channels that are consistently ``warm'' but not
energetic enough to disrupt triggering.  The energy thresholds used in
these definitions were determined from simulated QCD multi-jet samples and
CRAFT data, and were chosen so that physics signals do not cause the
channels to be mis-tagged as problematic.  The thresholds will be re-tuned and adjusted to appropriate values for each subdetector (HB, HE, HO and HF) on the basis of LHC collision data.

An HCAL channel is considered dead if no correctly-formatted data are
found from that channel in a sample of at least 10\,000 events, and if
this condition persists for at least 5\% of the events collected
in a given data-taking period.  A channel is also considered dead if it consistently fails to exceed a
given threshold energy.  The first test provides a check that
data are being received from each channel, and the second provides an
\emph{in situ} check of the energy reconstruction for that data.  As
with the hot channel algorithm, the dead channel thresholds are
user-controlled, and will be re-tuned on collision data.

Noise flags may be used to identify problematic channels or as a means to
reject entire events as noisy.  Two additional algorithms for providing rejection of noise events have been
developed.  These algorithms, after optimization and performance studies with first collision data, can be applied at the HLT to either flag events as potential noise or to reject them
completely.  The algorithms are: 

\begin{itemize}
\item{Rejection Algorithm I:} rejects any event containing a channel
with a hit multiplicity or charge ratio flag, as described above;
\item{Rejection Algorithm II:}  rejects events in which at least one RBX contains more than 8 channels with 0 ADC counts, and also rejects events in which any channel with energy greater than $25$~GeV has less than 70\% of its total charge contained within the signal pulse peak or falls outside a 13~ns window around the trigger;
\item{Rejection Algorithm III:} a more stringent version of Rejection
Algorithm II, requiring any channel with energy greater than
$25$~GeV to have 80\% of its total charge within its signal
peak, and to lie within a 9~ns window around the trigger.  Additionally,
no HPDs may contain more than 16 channels with energy above 1.5 GeV.
\end{itemize}

%% file: Performance.tex
The channel-flagging algorithms described in the previous section were tested on CRAFT
and single beam data taken in the fall of 2008.  They were also
applied to samples of simulated QCD multi-jets, $t\bar{t}$, and 1~TeV $Z'$ events at a
collision energy of 10~TeV.  The tests on simulated samples serve
as a cross-check of the algorithms, to ensure that expected physics
signals are not falsely tagged as noise.

\begin{figure}[!Hhtb]
  \centering
  \includegraphics[scale=.5]{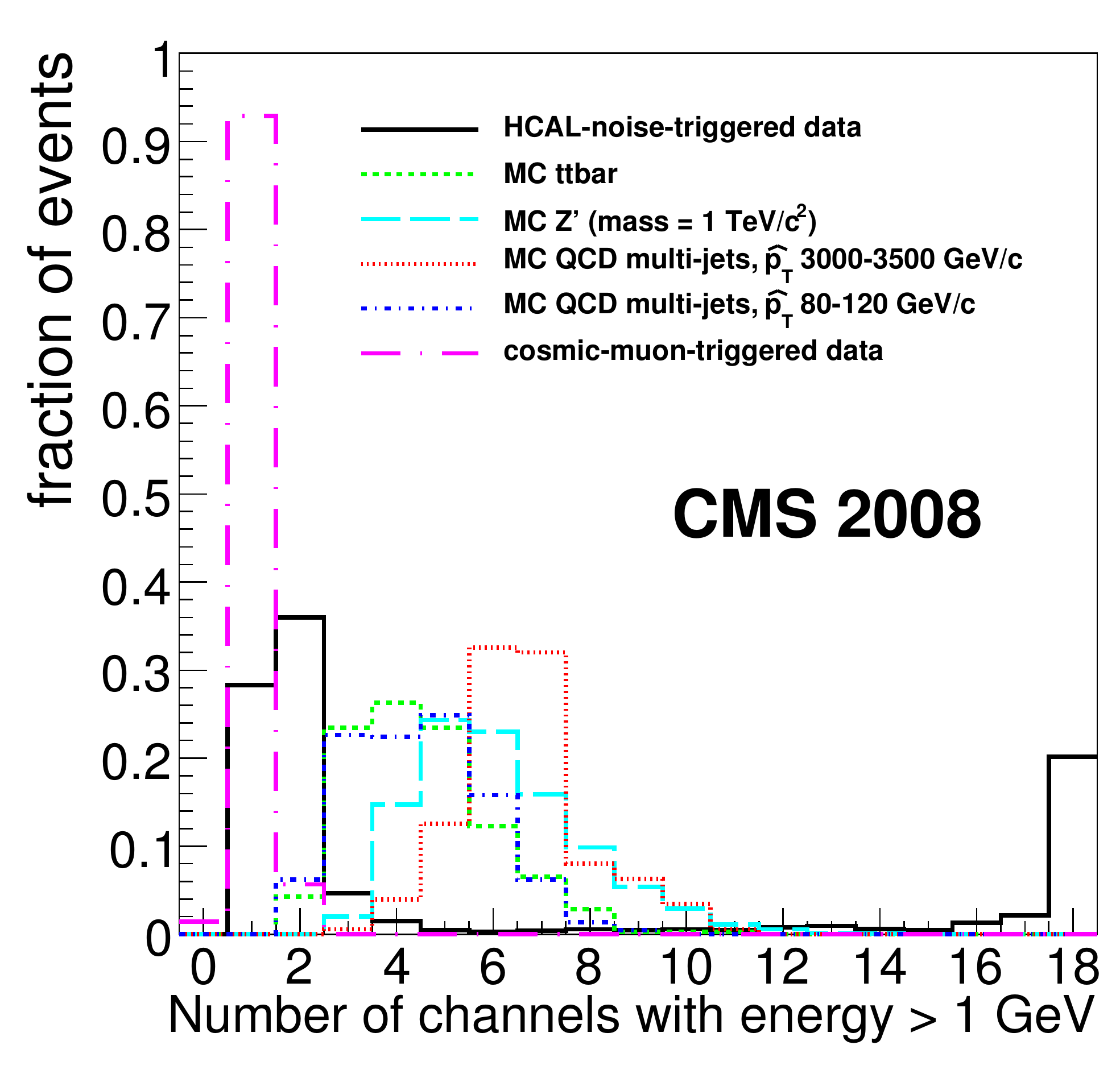}
  \caption{Number of HPD channels with energy $>$~1 GeV for the
  highest-multiplicity HPD in an event, for CRUZET noise and cosmic ray data, as well as simulated (MC) events.}
  \label{fig:HPDmultresults}
\end{figure}

\begin{figure}[!Hhtb]
\centering
    \includegraphics[scale=.5]{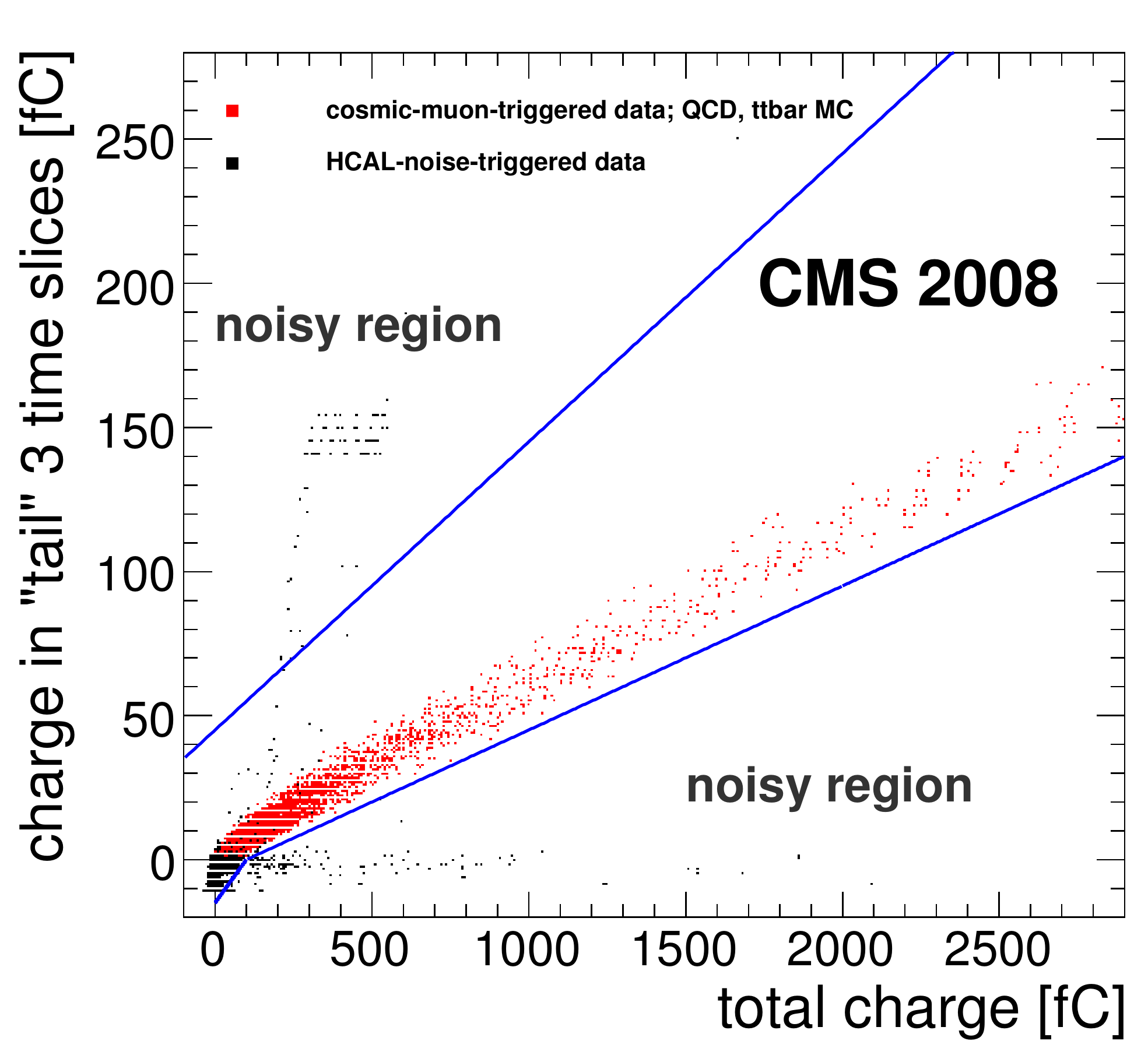}
    \caption{Sum of charges in the tail of the ADC pulse vs.\ sum of
    charges over the entire pulse.  The lines indicate the noisy regions; channels falling within these regions are flagged as noisy.}
    \label{fig:HPDshaperesults}
\end{figure}

The channel-flagging algorithms that detect problematic HPD output were applied to
data from special CRUZET noise data triggered on events where two
neighboring HCAL towers contained at least 10~GeV of energy.  The
algorithms were also applied to QCD multi-jet, $t\bar{t}$,
and $Z'\to t\bar{t}$ simulations to evaluate their effects on real
physics signals.  The distribution of HPD hit multiplicities is shown in
Fig. \ref{fig:HPDmultresults}.  HPD noise, in which all 18 channels within an HPD report energy deposits greater than 1~GeV, is evident in
the noise data, while simulated events and data triggered
on cosmic ray muons show lower multiplicities.
Figure \ref{fig:HPDshaperesults} shows the distribution of ADC pulse
tails vs. total collected charge, and illustrates the difference in
tail distributions between muon-triggered and noise-triggered data. 

The noise requirements as defined in Fig. \ref{fig:HPDshaperesults}
are loose enough not to falsely tag any physics signal as noise.  These loose cuts allow for minor differences in ADC pulse shape between real and simulated data.  In tests over a few thousand simulated events, no channels were flagged
as noisy by the algorithms.  The algorithm was also applied to a
subset of CRUZET events that passed a special noise trigger requiring
at least 40~GeV of energy deposited within a single RBX.  In these
HCAL-triggered noise data, 94\% of all events contained at least one
channel flagged as noisy.

\begin{figure}[!Hhtb]
  \centering
  \includegraphics[scale=.55]{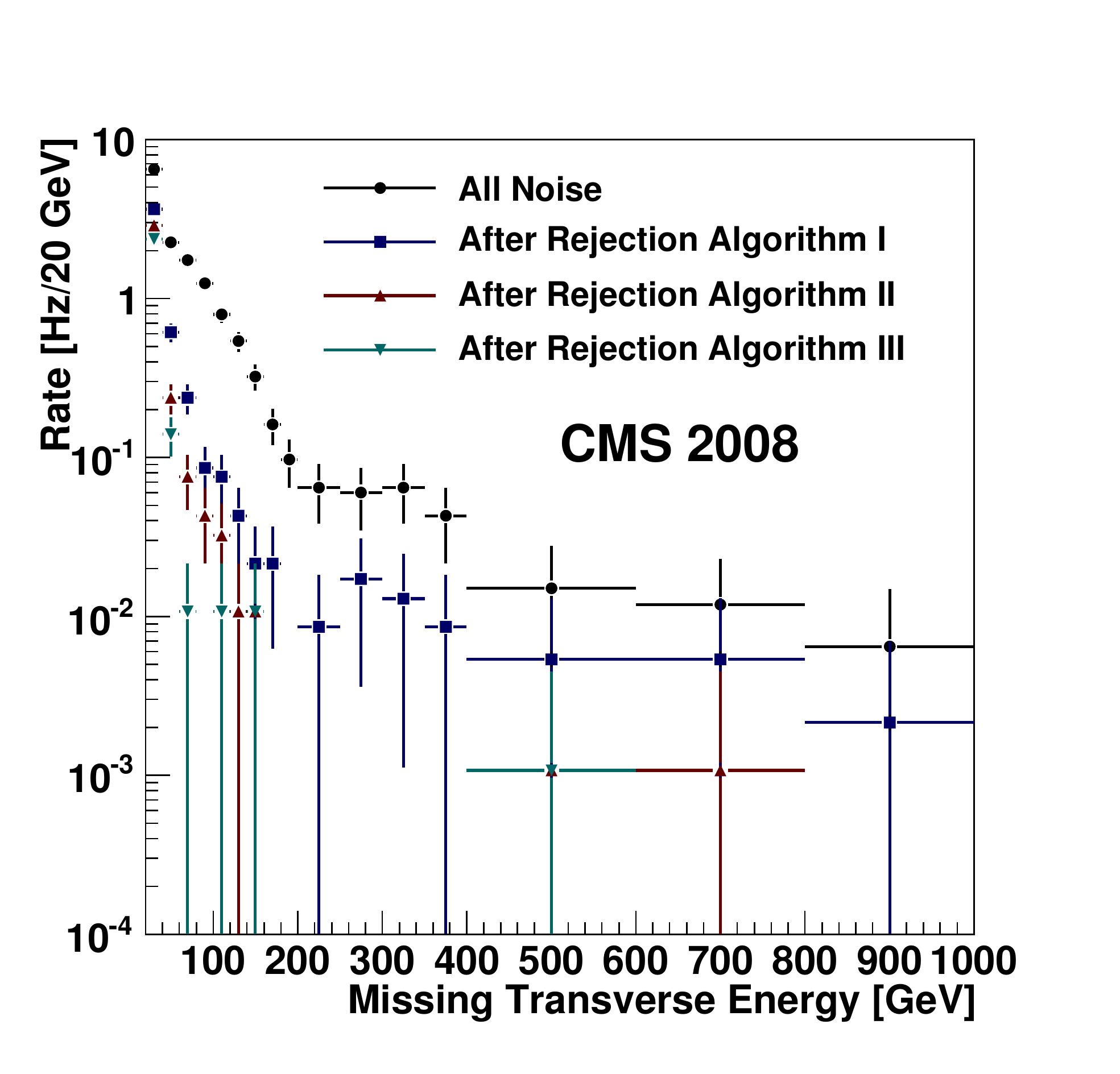}
  \caption{HCAL noise rates vs.\ MET in CRAFT
  data, before and after the application of noise rejection algorithms.  Rates were determined from events passing a 6~GeV jet trigger, and with at least 40~GeV of measured MET.}

  \label{fig:flags}
\end{figure}

The rate of fake MET signals in CRAFT data
before and after the application of HB/HE noise reduction algorithm is shown 
in Fig. \ref{fig:flags}.  The application of Rejection Algorithm I can reduce fake noise
signals by a factor of 6.5~(10) for events with MET greater
than 45~(100)~GeV.  Rejection Algorithms II and III are
substantially more efficient than Algorithm I at removing noise,
but they also reject a small fraction of physics signal.  Algorithm II falsely rejects $\sim0.4\%$
of all simulated $t\bar{t}$ events, while Algorithm III rejects $\sim3\%$ of such events.  
  
\begin{figure}[!Hhtb]
  \centering
 \includegraphics[scale=.55]{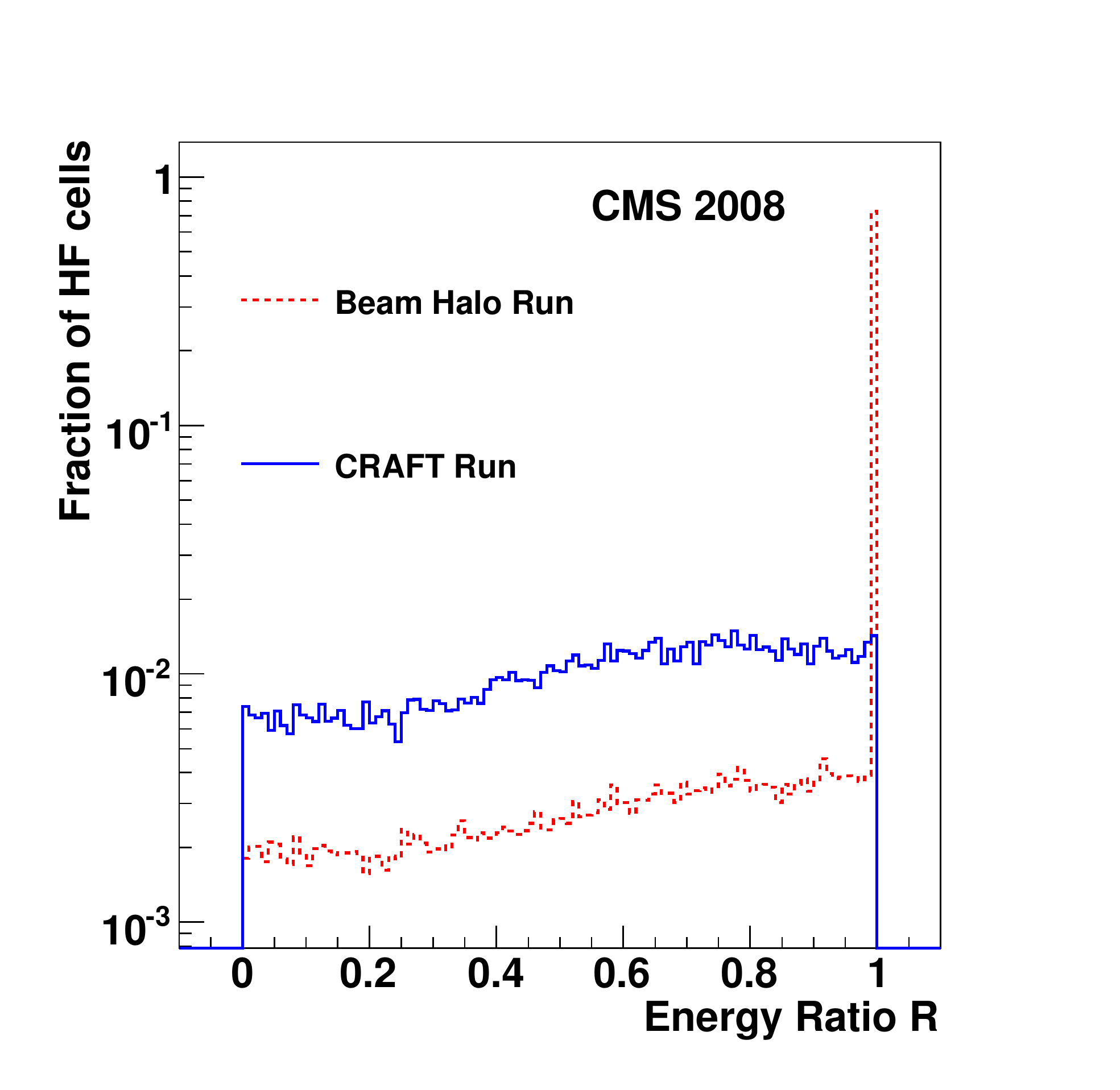}
 \caption{Ratios of energy differences to energy sums of adjacent long
 and short HF fiber channels in CRAFT and LHC beam halo data.
 Long and short channels with ratios greater than 0.99 are flagged as noisy.}
  \label{fig:HF_resultplot}
\end{figure}

Testing of the HF noise-tagging algorithm was performed on CRAFT data and LHC beam halo data.  The CRAFT results were collected from events passing a muon trigger, while beam halo events were required to pass either an HF calorimeter trigger or a muon system halo trigger.  High muon rates in the beam halo
data resulted in large numbers of PMT window hits, as shown in Fig. \ref{fig:HF_resultplot}.  Tests of the saturation algorithm showed no saturated HCAL
channels in either CRAFT cosmics data or simulated $t\bar{t}$
samples.  Beam halo events also resulted in no saturated channels,
while splash events produced an average of 20 saturated HF channels.
The HCAL timing synchronization algorithm was also used to identify synchronization errors in CRAFT data.  The resultant post-CRAFT upgrades to the HCAL
timing have lowered the rate of such synchronization errors to less than 0.1\%.

Searches for persistent problems during CRAFT identified a single hot
channel in HF.  This channel was repaired, and there are currently no hot channels within HCAL.  No
dead channels were observed in HB, HE, or HF.  A total of five
dead channels was found in the region of HO that is being used in
event reconstruction.  These channels are not accessible for
repair, and have been recorded as dead in the status database.  The
total fraction of dead channels within the HCAL is less than 0.25\%.

%% file: CFT-09-019-authorlist.tex
\textbf{Yerevan Physics Institute,  Yerevan,  Armenia}\\*[0pt]
S.~Chatrchyan, V.~Khachatryan, A.M.~Sirunyan
\vskip\cmsinstskip
\textbf{Institut f\"{u}r Hochenergiephysik der OeAW,  Wien,  Austria}\\*[0pt]
W.~Adam, B.~Arnold, H.~Bergauer, T.~Bergauer, M.~Dragicevic, M.~Eichberger, J.~Er\"{o}
% ...
\vskip\cmsinstskip
% ...


%% file: CFT-09-019_temp.bbl
\providecommand{\href}[2]{#2}\begingroup\raggedright\begin{thebibliography}{1}

\bibitem{CMSjinst}
{CMS Collaboration}, ``{The CMS experiment at the CERN LHC}'', {\em JINST} {\bf
  3} (2008)
S08004.
%%CITATION = JINST,0803,S08004;%%.
  \href{http://dx.doi.org/10.1088/1748-0221/3/08/S08004}{{\tt
  doi:10.1088/1748-0221/3/08/S08004}}.

\bibitem{LHC}
{L. Evans and P. Bryant (eds.)}, ``LHC Machine'', {\em JINST} {\bf 3} (2008)
  S08001.

\bibitem{HcalNoise}
{CMS Collaboration}, ``Performance of the CMS Hadron Calorimeter with Cosmic
  Ray and Accelerator Produced Muons'', {\em Submitted to JINST} (2009).

\bibitem{HBdesign}
S.~Abdullin {et~al.}, ``{Design, performance, and calibration of CMS
  hadron-barrel calorimeter wedges}'', {\em Eur. Phys. J.} {\bf C55} (2008)
159--171.
%%CITATION = EPHJA,C55,159;%%.
  \href{http://dx.doi.org/10.1140/epjc/s10052-008-0573-y}{{\tt
  doi:10.1140/epjc/s10052-008-0573-y}}.

\bibitem{HFdesign}
S.~Abdullin {et~al.}, ``{Design, performance, and calibration of the CMS
  forward calorimeter wedges}'', {\em Eur. Phys. J.} {\bf C53} (2008)
139--166.
%%CITATION = EPHJA,C53,139;%%.
  \href{http://dx.doi.org/10.1140/epjc/s10052-007-0459-4}{{\tt
  doi:10.1140/epjc/s10052-007-0459-4}}.

\bibitem{CONDDB}
M.~De~Gruttola, S.~di~Guida, F.~Glege{ et~al.}, ``First experience in operating
  the population of the condition database for the CMS experiment'', {\em CMS
  Conference Report} {\bf
  \href{http://cms.cern.ch/iCMS/jsp/openfile.jsp?type=CR&year=2009&files=CR200%
9_103.pdf}{2009/103}} (2009).

\bibitem{HPDNOISE}
R.~Korzekwa, F.~M. Lehr, H.~G. Krompholz{ et~al.}, ``The influence of magnetic
  fields on dielectric surface flashover'', {\em IEEE Transactions on Electron
  Devices} {\bf 38} (April, 1991) 745--749.

\bibitem{HCAL_PHIL}
{CMS Collaboration}, ``Performance of CMS Hadron Calorimeter Timing and
  Synchronization Using Cosmic Ray and LHC Beam Data'', {\em Submitted to
  JINST} (2009).

\end{thebibliography}\endgroup
